\documentclass[preprint]{cimento}

\usepackage{graphicx} 

\title{LHC and ILC Data and the Early Universe Properties}
\author{A.~Arbey\from{ins:CRAL},
F.~Mahmoudi\from{ins:LPC}}

\instlist{
\inst{ins:CRAL} Universit\'e de Lyon, France; Universit\'e Lyon 1, F--69622; CRAL, Observatoire de Lyon, F--69561 Saint-Genis-Laval; CNRS, UMR 5574; ENS de Lyon, France.
\inst{ins:LPC} Clermont Universit\'e, Universit\'e Blaise Pascal, CNRS/IN2P3,\\ LPC, BP 10448, 63000 Clermont-Ferrand, France.
}

\PACSes{
\PACSit{11.30.Pb}{Supersymmetry}
\PACSit{95.35.+d}{Dark matter}
\PACSit{98.80.Cq}{Particle-theory and field-theory models of the early Universe}
}

\begin{document}

\maketitle

\begin{abstract}
With the start-up of the LHC, we can hope to find evidences for new physics beyond the Standard Model, and particle candidates for dark matter. Determining the parameters of the full underlying theory will be a long process requiring the combination of LHC and ILC data, flavor physics constraints, and cosmological observations. However, the Very Early Universe properties, from which the relic particles originate, are poorly known, and the relic density calculation can be easily falsified by hidden processes. We consider supersymmetry and show that determining the underlying particle physics parameters will help understanding the Very Early Universe properties.
\end{abstract}

\section{Introduction}

The present and future high energy colliders will hopefully allow the discovery of new particles. Many new physics models beyond the Standard Model propose particle candidates for dark matter, within the reach of the LHC and ILC. Should a stable, neutral and weakly interacting new particle be found, it would be a candidate for dark matter. Through particle physics computation of annihilation and co-annihiliation diagrams, it is possible to compute the dark matter relic density \cite{relic_calculation}. This relic density is then often compared to the dark matter density deduced from cosmological observations in order to constrain new physics parameters. If the computed relic density is compatible with the cosmologically inferred dark matter density, the cosmological model will be reinforced by this new particle discovery. However, in case of disagreements, two possible paths can open: either the new physics particle model is not correctly designed, or the cosmological model is missing ingredients, such as quintessence \cite{quintessence} or reheating \cite{reheating}. In \cite{arbey,arbey2}, we showed that such dark components can modify the computed relic density by several orders of magnitudes, and we introduced parametrizations to characterize the Very Early Universe properties. Using this kind of parametrizations, and combining with particle physics data, it will therefore be possible to determine the Very Early Universe properties beyond the standard cosmological scenario. In the following we will consider Supersymmetry (SUSY) as an example and discuss the use of relic density to constrain cosmological properties.

\section{Relic density calculation}
The density number of supersymmetric particles is determined by the Boltzmann equation and takes the form:
\begin{equation}
\frac{dn}{dt} = - 3 H n - \langle \sigma v \rangle (n^2 - n^2_{eq}) \;,\label{boltzmann}
\end{equation}
where $n$ is the number density of supersymmetric particles, $\langle \sigma v \rangle$ is the thermally averaged annihilation cross-section, $H$ is the Hubble expansion rate and $n_{eq}$ is the supersymmetric particle equilibrium number density. The expansion rate $H$ is determined by the Friedmann equation:
\begin{equation}
 H^2=\frac{8 \pi G}{3} (\rho_{rad} + \rho_D)  \;.\label{friedmann}
\end{equation}
$\rho_{rad}$ is the radiation energy density, which is considered to be dominant before BBN in the standard cosmological model. $\rho_D$ is introduced in Eq.~(\ref{friedmann}) to parametrize the expansion rate modification \cite{arbey} and can be interpreted either as an additional energy density modifying the expansion ({\it e.g.} quintessence), or as an effective energy density which can account for other phenomena affecting the expansion rate ({\it e.g.} extra-dimensions).

The entropy evolution can also be altered beyond the standard cosmological model. In presence of entropy fluctuations we give the entropy evolution equation:
\begin{equation}
\frac{ds}{dt} = - 3 H s + \Sigma_D \label{entropy_evolution} \;,
\end{equation}
where $s$ is the total entropy density. $\Sigma_D$ in the above equation parametrizes effective entropy fluctuations due to unknown properties of the Early Universe, and is temperature-dependent.

The parameters $\rho_D$ and $\Sigma_D$ can be regarded as independent. Entropy and energy alterations are considered here as effective effects, and can be generated by curvature, phase transitions, extra-dimensions, or other phenomena in the Early Universe. In a specific physical scenario, these parameters can be related, as for example in reheating models \cite{reheating}. 

The radiation energy and entropy densities can be written as usual:
\begin{equation}
\rho_{rad}=g_{eff}(T) \frac{\pi^2}{30} T^4 \;, \qquad\qquad s_{rad} = h_{eff}(T) \frac{2\pi^2}{45} T^3 \;. \label{srad}
\end{equation}
We split the total entropy density into two parts: radiation entropy density and effective dark entropy density, $s \equiv s_{rad} + s_D$. Using Eq.~(\ref{entropy_evolution}) the relation between $s_D$ and $\Sigma_D$ can then be derived:
\begin{equation}
\Sigma_D = \sqrt{\frac{4 \pi^3 G}{5}} \sqrt{1 + \tilde{\rho}_D} T^2 \left[\sqrt{g_{eff}} s_D - \frac13  \frac{h_{eff}}{g_*^{1/2}} T \frac{ds_D}{dT}\right] \;,
\end{equation}
with
\begin{equation}
g_*^{1/2} = \frac{h_{eff}}{\sqrt{g_{eff}}}\left(1+\frac{T}{3 h_{eff}} \frac{dh_{eff}}{dT}\right) \;.
\end{equation}
Following the standard relic density calculation method \cite{relic_calculation}, $Y \equiv n/s$ is introduced, and Eq.~(\ref{boltzmann}) yields
\begin{equation}
 \frac{dY}{dx}= - \frac{m_{lsp}}{x^2} \sqrt{\frac{\pi}{45 G}} g_*^{1/2} \left( \frac{1 + \tilde{s}_D}{\sqrt{1+\tilde{\rho}_D}} \right) \left[\langle \sigma v \rangle (Y^2 - Y^2_{eq}) + \frac{Y \Sigma_D}{\left(h_{eff}(T) \frac{2\pi^2}{45} T^3\right)^2 (1+\tilde{s}_D)^2} \right] \;, \label{final}
\end{equation}
where $x=m_{lsp}/T$, $m_{lsp}$ is the mass of the lightest supersymmetric relic particle, and
\begin{equation}
 \tilde{s}_D \equiv \frac{s_D}{h_{eff}(T) \frac{2\pi^2}{45} T^3}\;, \qquad\qquad \tilde{\rho}_D \equiv \frac{\rho_D}{g_{eff} \frac{\pi^2}{30} T^4}\;,
\end{equation}
and
\begin{equation}
 Y_{eq} = \frac{45}{4 \pi^4 T^2} h_{eff} \frac{1}{(1+\tilde{s}_D)} \sum_i g_i m_i^2 K_2\left(\frac{m_i}{T}\right) \;,
\end{equation}
with $i$ running over all supersymmetric particles of mass $m_i$ and with $g_i$ degrees of freedom. Integrating Eq. (\ref{final}), the relic density can then be calculated using:
\begin{equation}
 \Omega h^2 = \frac{m_{lsp} s_0 Y_0 h^2}{\rho_c^0} = 2.755 \times 10^8 Y_0 m_{lsp}/\mbox{GeV} \;, \label{omegah2}
\end{equation}
where the subscript 0 refers to the present value of the parameters. In the limit where $\rho_D = s_D = \Sigma_D = 0$, standard relations are retrieved. Using Eqs. (\ref{boltzmann}-\ref{omegah2}) the relic density in presence of a modified expansion rate and of entropy fluctuations  can be computed provided $\rho_D$ and $s_D$ are specified.
For $\rho_D$ we follow the parametrization introduced in Ref. \cite{arbey}:
\begin{equation}
 \rho_D =  \kappa_\rho \rho_{rad}(T_{BBN}) \bigl(T/T_{BBN}\bigr)^{n_\rho} \;, \label{rhoD}
\end{equation}
where $T_{BBN}$ is the BBN temperature. Different values of $n_\rho$ leads to different behaviors of the effective density. For example, $n_\rho=4$ corresponds to a radiation behavior, $n_\rho=6$ to a quintessence behavior, and $n_\rho>6$ to the behavior of a decaying scalar field. $\kappa_\rho$ is the ratio of the effective energy density to the radiation energy density at BBN time and can be negative. The role of $\rho_D$ is to increase the expansion rate for $\rho_D > 0$, leading to an early decoupling and a higher relic density, or to decrease it for $\rho_D < 0$, leading to a late decoupling and to a smaller relic density. To model the entropy perturbations, we follow the parametrization introduced in Ref. \cite{arbey2}:
\begin{equation}
 s_D =  \kappa_s s_{rad}(T_{BBN}) \bigl(T/T_{BBN}\bigr)^{n_s} \;. \label{sD}
\end{equation}
This parametrization finds its roots in the first law of thermodynamics, where energy and entropy are directly related and therefore the entropy parametrization can be similar to the energy parametrization. As for the energy density, different values of $n_s$ lead to different behaviors of the entropy density: $n_s = 3$ corresponds to a radiation behavior, $n_s = 4$ appears in dark energy models, $n_s \sim 1$ in reheating models, and other values can be generated by curvature, scalar fields or extra-dimension effects. $\kappa_s$ is the ratio of the effective entropy density to the radiation entropy density at BBN time and can be negative. The role of $s_D$ is to increase the temperature at which the radiation dominates for $s_D > 0$, leading to a decreased relic density, or to decrease this temperature for $s_D < 0$, increasing the relic density. Constraints on the cosmological entropy in reheating models have been derived in \cite{entropy_constraints}. 

\section{SUSY constraints}
We now consider the effects of the parametrizations described in the previous section on the supersymmetric constraints. Using the latest WMAP data \cite{WMAP5} with an additional 10\% theoretical uncertainty on the relic density calculation, we give the following favored interval at 95\% C.L.:
\begin{equation}
 0.088 < \Omega_{DM} h^2 < 0.123 \;.\label{wmap}
\end{equation}
The older dark matter interval is also considered:
\begin{equation}
 0.1 < \Omega_{DM} h^2 < 0.3 \;.\label{old}
\end{equation}
One million random SUSY points in the NUHM parameter plane ($\mu$,$m_A$) with $m_0=m_{1/2}=1$ TeV, $A_0=0$, $\tan\beta=40$ are generated using SOFTSUSY v2.0.18 \cite{softsusy}, and for each point we compute flavor physics observables, direct limits and the relic density with SuperIso Relic v2.7 \cite{superiso,superiso_relic}. %
\begin{figure}[!t]
\begin{center}
\includegraphics[width=6.cm]{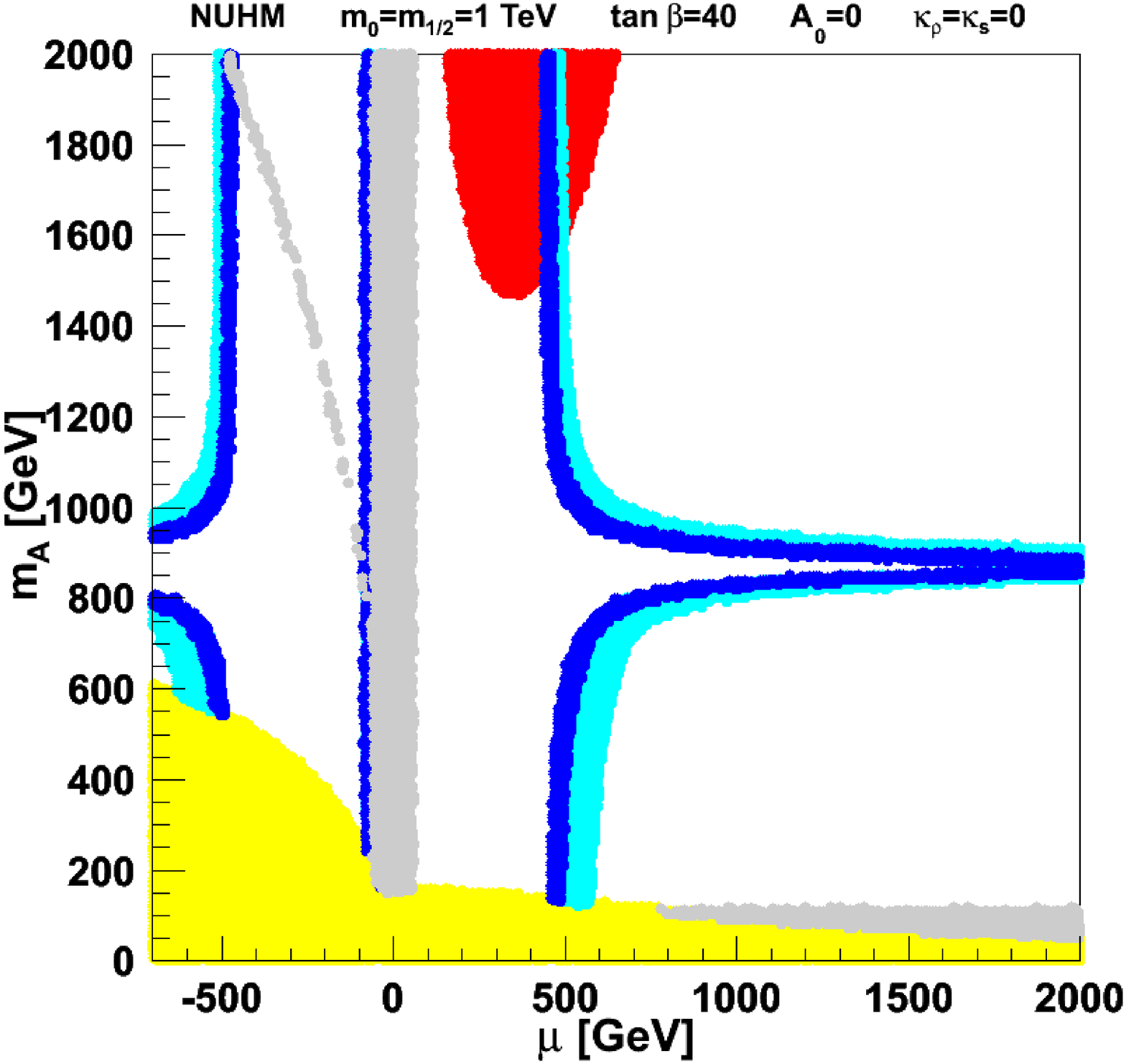}\hspace*{0.2cm}\includegraphics[width=6.cm]{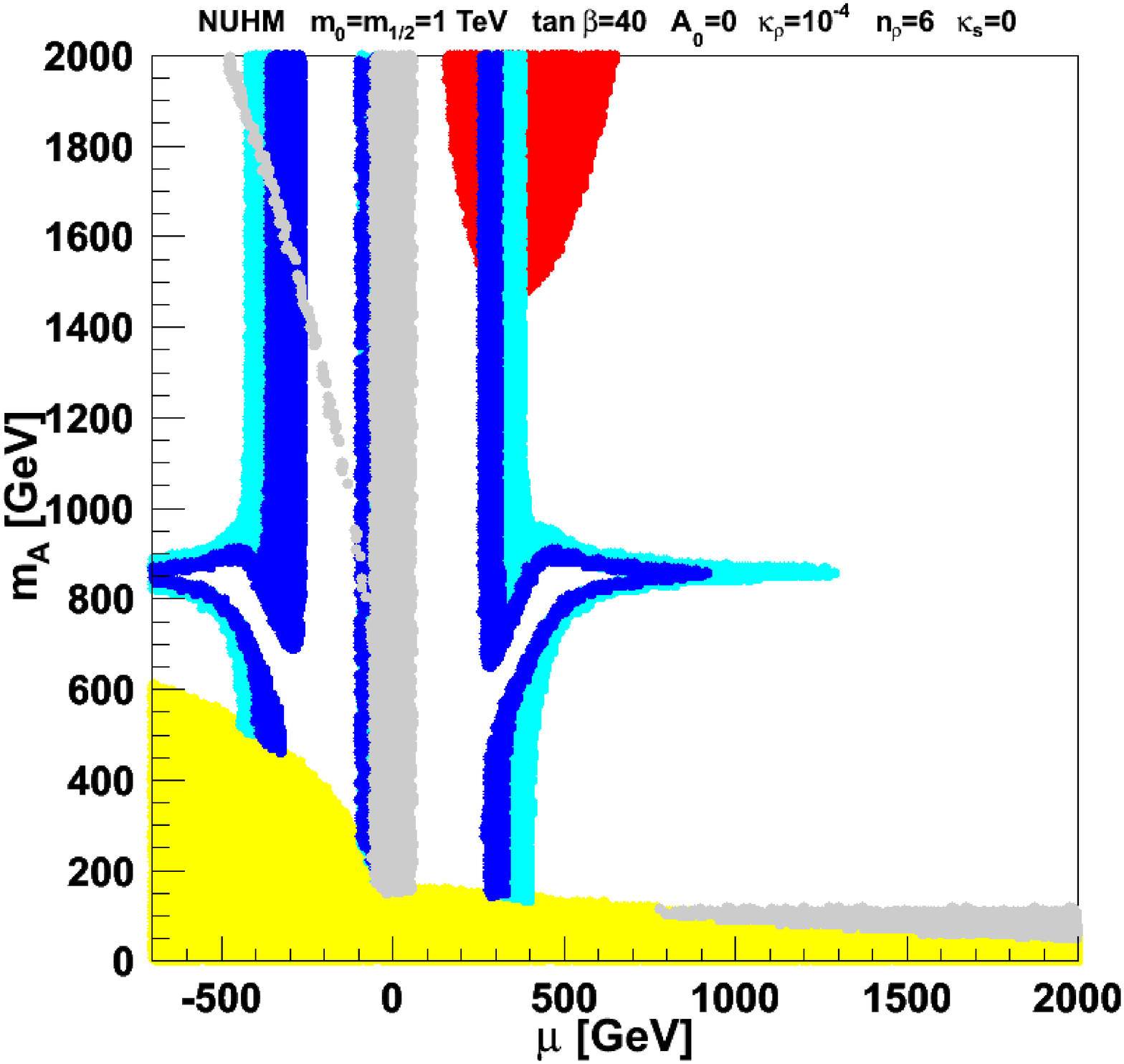}\\[0.1cm]
\includegraphics[width=6.cm]{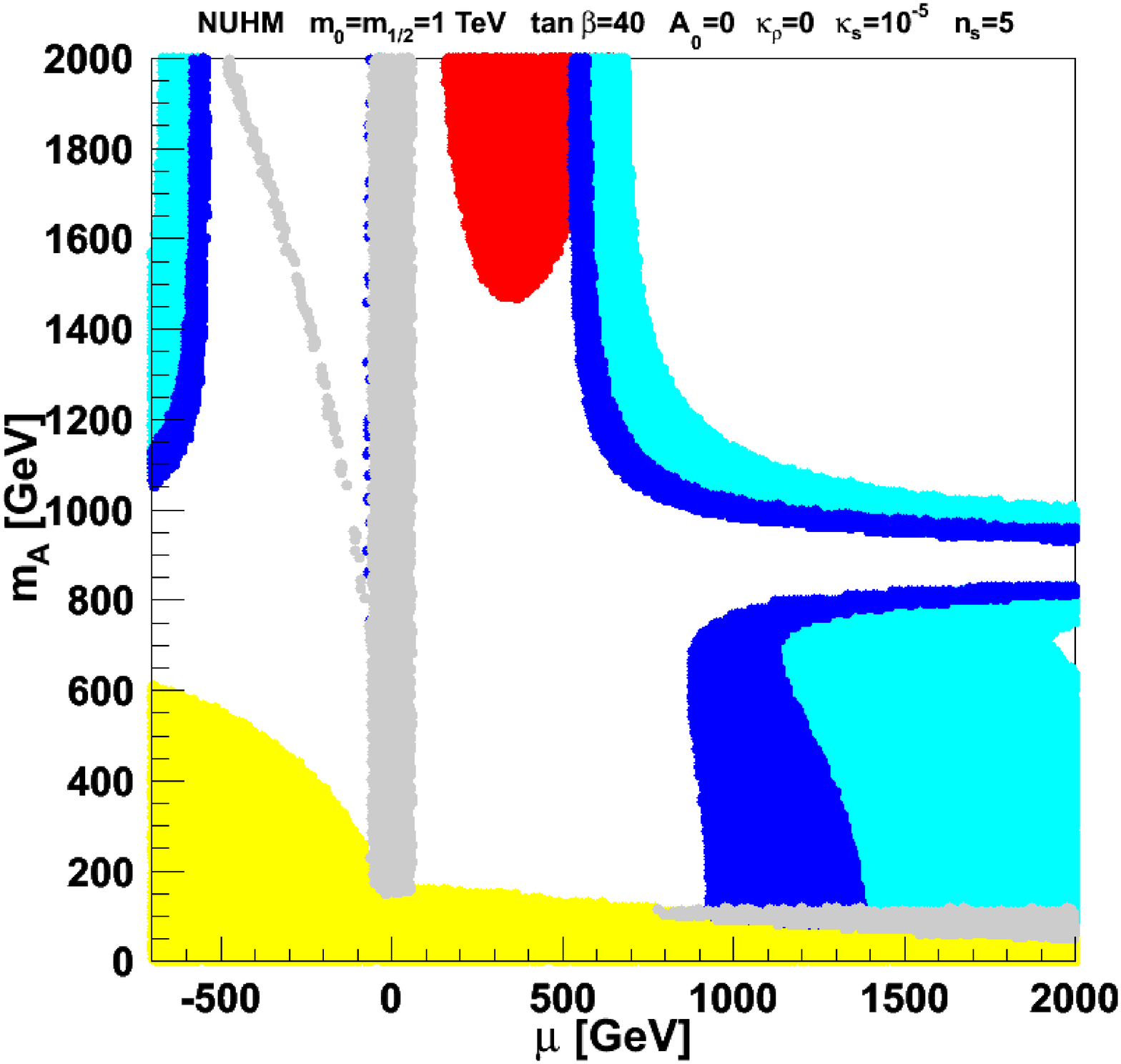}\hspace*{0.2cm}\includegraphics[width=6.cm]{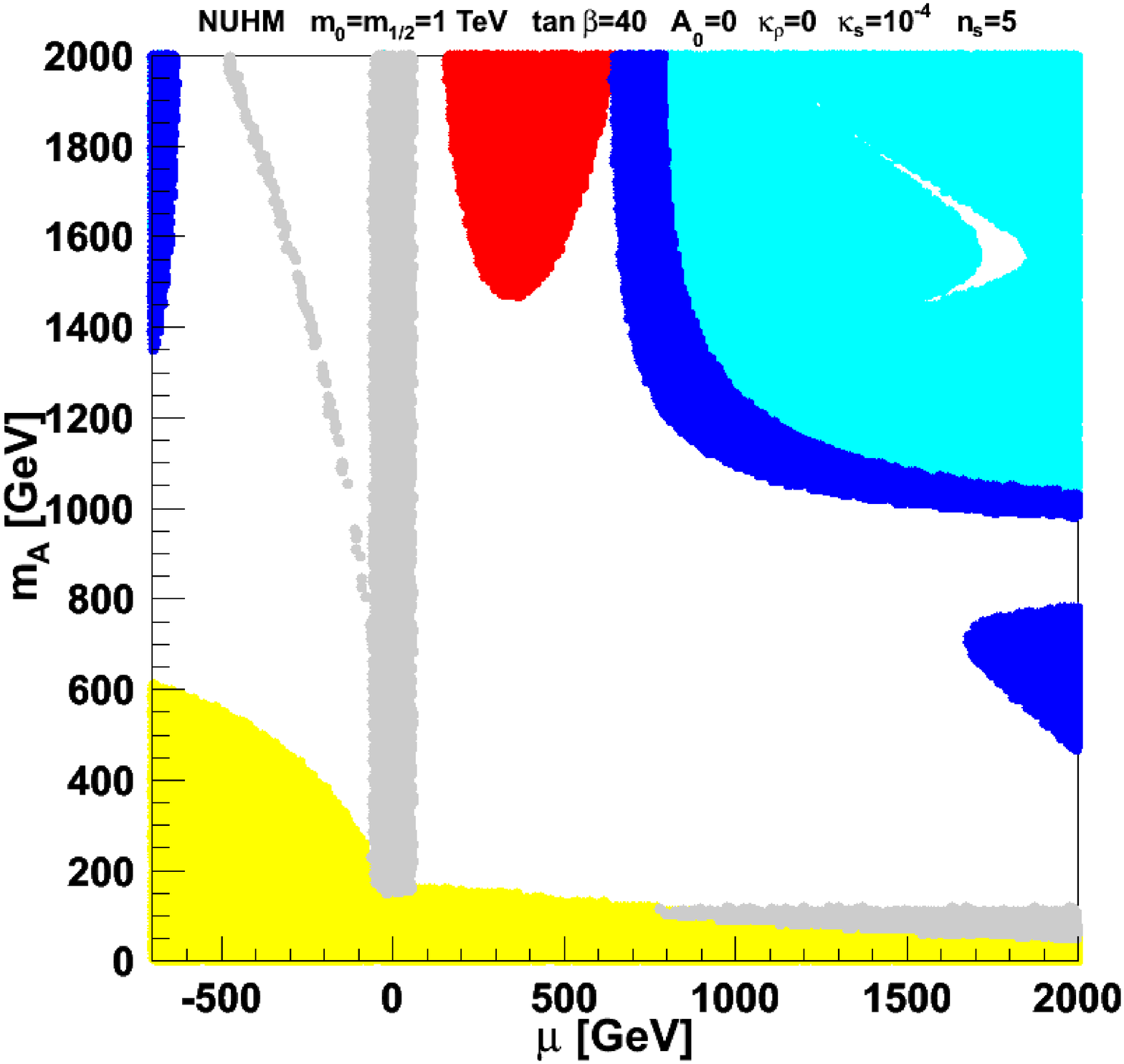}
\end{center}
\caption{Constraints on the NUHM parameter plane ($\mu$,$m_A$), in the standard cosmological model (top left), in presence of a tiny energy overdensity with $\kappa_\rho=10^{-4}$ and $n_\rho=6$ (top right), and of an entropy overdensity with $\kappa_s=10^{-5}$ and $n_s=5$ (bottom left), with $\kappa_s=10^{-4}$ and $n_s=5$ (bottom right). The red points are excluded by the isospin asymmetry of $B \to K^* \gamma$, the gray points by direct collider limits, the yellow zones involve tachyonic particles, and the dark and light blue strips are \underline{favored} by the dark matter constraints of Eqs. (\ref{wmap}) and (\ref{old}) respectively.\label{fig3}}
\end{figure}%
In Fig.~\ref{fig3}, the effects of the cosmological models on the relic density constraints are demonstrated. The first plot is given as a reference for the standard cosmological model, showing the tiny strips corresponding to the regions favored by the relic density constraint. In the second plot, generated in a Universe with an additional energy density with $\kappa_\rho=10^{-4}$ and $n_\rho=6$, the relic density favored strips are reduced, since the calculated relic densities are decreased in comparison to the relic densities computed in the standard scenario. 
The next plots demonstrate the influence of an additional entropy density compatible with BBN constraints. The favored strips are this time enlarged and moved towards the outside of the plot. This effect is due to a decrease in the relic density.
These figures show that a modification in the cosmological scenario can completely modify the calculated relic density and lead to different shapes of the favored parameter regions.

\section{Inverse Problem}
\begin{figure}[!t]
\begin{center}
\includegraphics[width=6.cm]{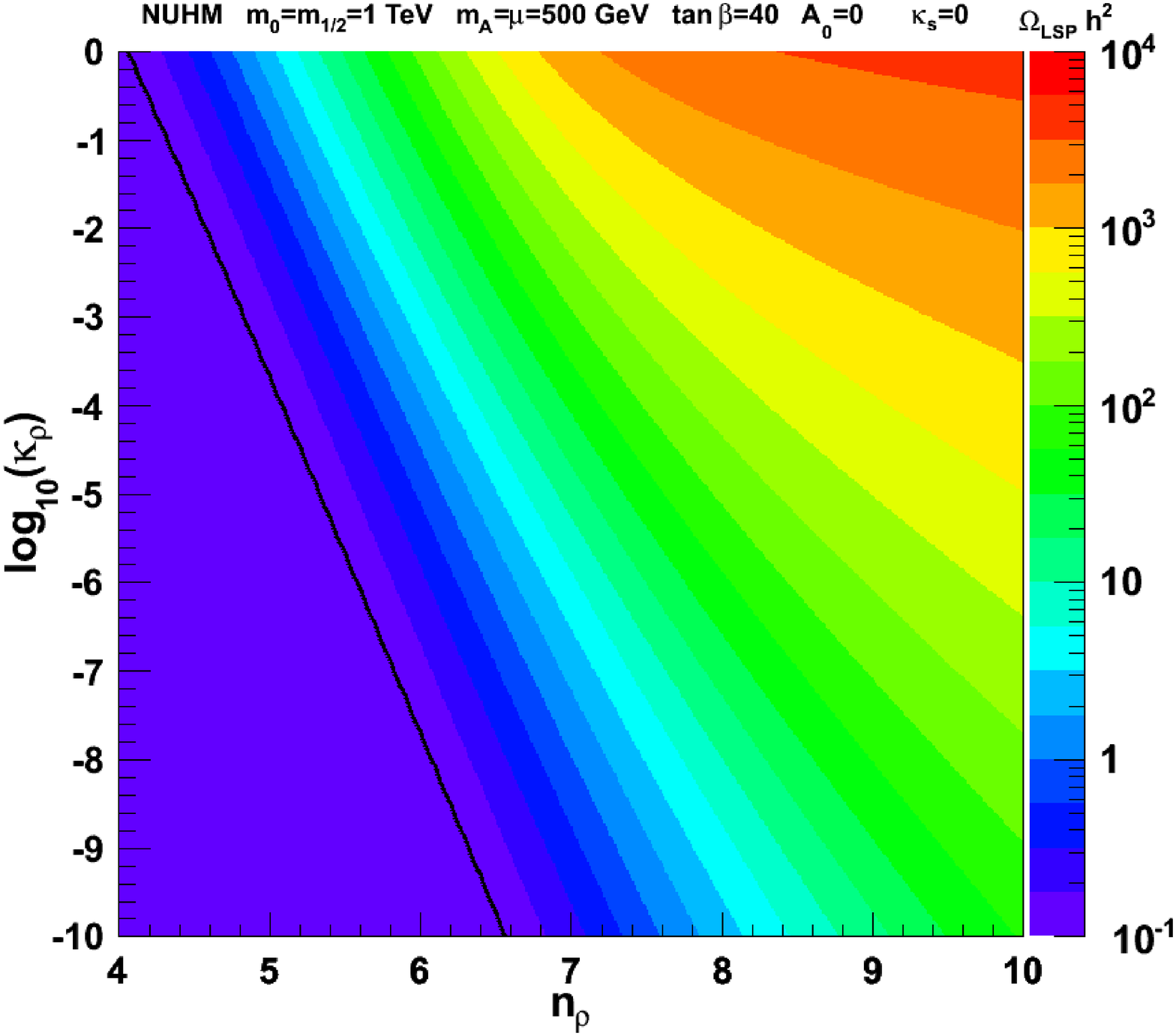}\hspace*{0.2cm}\includegraphics[width=6.cm]{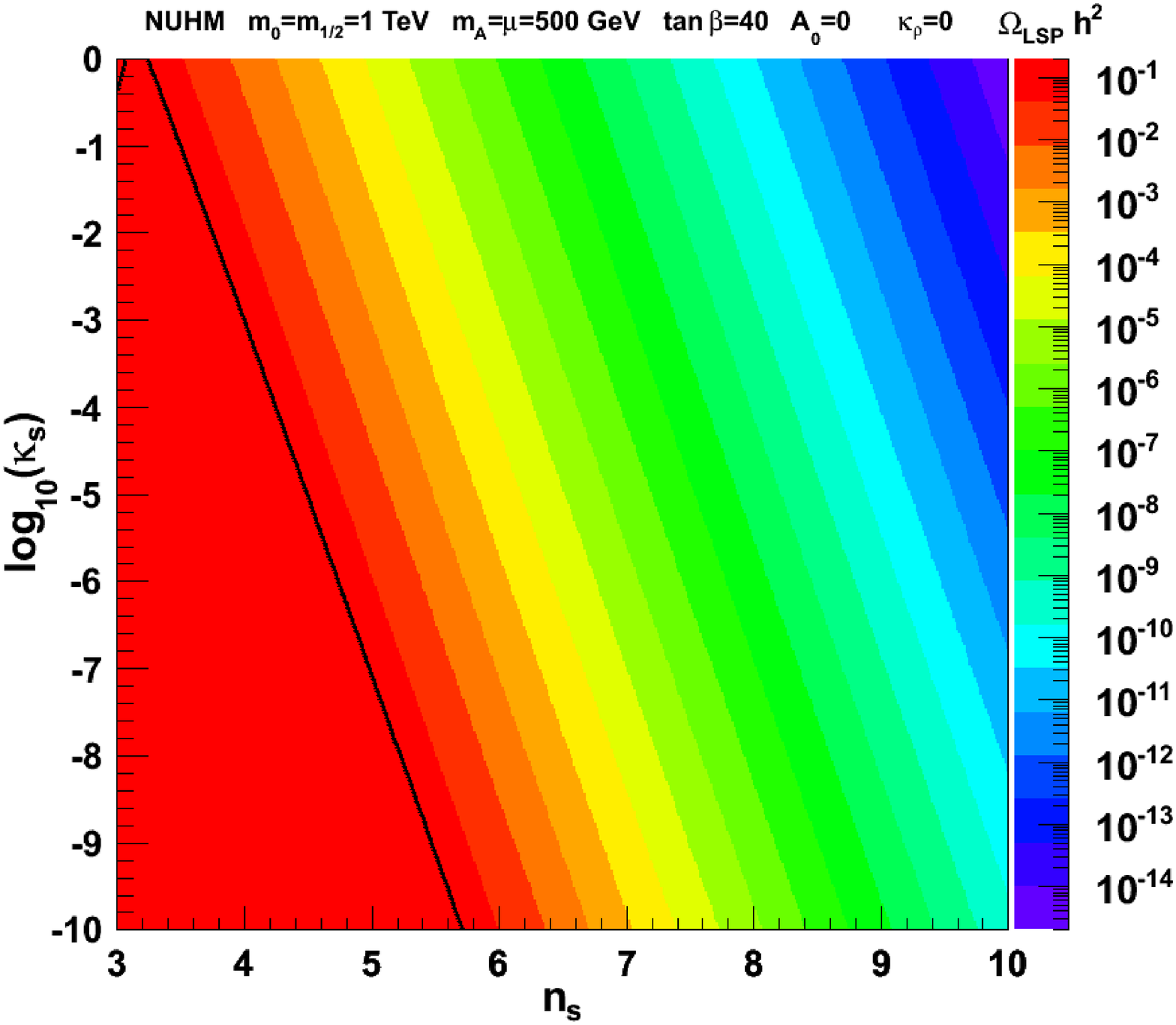}
\end{center}
\caption{Influence of the presence of an effective energy density with $n_\rho=6$ (left), and an effective entropy with $n_s=5$ (right). The colors correspond to different values of $\Omega h^2$. The black lines delimit the regions favored by WMAP. The favored zones are the lower left corners.\label{fig2}}
\end{figure}%

The determination of the supersymmetric parameters using particle physics observables can on the other hand give access to global properties of the relic particle decoupling period \cite{inverse}.
Let us consider the NUHM example point ($m_0=m_{1/2}=1$ TeV, $m_A=\mu=500$ GeV, $A_0=0$, $\tan\beta=40$), which gives a relic density of $\Omega h^2 \approx 0.11$ in the cosmological standard model, compatible with WMAP results. The effects due to the presence of effective energy or entropy in the Early Universe are presented in Fig. \ref{fig2}: the first plot shows the influence of an additional effective density on the computed relic density. We note that when $\kappa_\rho$ and $n_\rho$ increase, the relic density increases up to a factor of $10^5$. The second plot illustrates the effect of an additional entropy density, in absence of additional energy density. Here when $\kappa_s$ and $n_s$ increase, the relic density is strongly decreased down to a factor of $10^{-14}$. The dark lines delimit the zones which are compatible with WMAP data. The determination of the NUHM parameters provides therefore interesting constraints on the cosmological properties of the Early Universe.

It is important to point out that all the cosmological scenarios previously described are equivalent from the point of view of the cosmological observations: there is no way to distinguish between them with the current cosmological data. 

\section{Conclusion}
We discussed the use of the LHC and ILC data, as well as flavor physics constraints, together with the dark matter relic density evaluation, to explore the properties of the Very Early Universe. In particular, we provided parametrizations of the entropy content and the expansion rate for this purpose. We have shown that a disagreement between the computed relic density and the observed dark matter density may be a sign of deviation from the standard cosmological model. Should high energy colliders discover candidates for dark matter, such studies of the Early Universe properties will become possible.

\end{document}